\documentclass[%
 reprint,
superscriptaddress,
 amsmath,amssymb,
 aps,
prl,
]{revtex4-1}

\usepackage{graphicx}
\usepackage{dcolumn}
\usepackage{bm}
\usepackage{braket}
\usepackage{xcolor}
\usepackage{float}
\usepackage{subfiles}
\usepackage{comment}
\usepackage[export]{adjustbox}
\usepackage{hyperref} 

\hypersetup{
    colorlinks=true,
    linkcolor=blue,
    citecolor=blue,
    filecolor=blue,
    urlcolor=blue
}

\begin{document}

\preprint{APS/123-QED}

\title{Primary quantum thermometry of mm-wave blackbody radiation  via \\induced state transfer in Rydberg states of cold atoms}

\author{Noah~Schlossberger}
\email{noah.schlossberger@nist.gov}
\affiliation{National Institute of Standards and Technology, Boulder, Colorado 80305, USA}

\author{Andrew~P.~Rotunno}
\affiliation{National Institute of Standards and Technology, Boulder, Colorado 80305, USA}

\author{Stephen~P.~Eckel}
\affiliation{National Institute of Standards and Technology, Gaithersburg, MD 20899 USA}

\author{Eric~B.~Norrgard}
\affiliation{National Institute of Standards and Technology, Gaithersburg, MD 20899 USA}

\author{Dixith~Manchaiah}
\affiliation{Associate of the National Institute of Standards and Technology, Boulder, Colorado 80305, USA}
\affiliation{Department of Physics, University of Colorado, Boulder, Colorado 80309, USA}

\author{Nikunjkumar~Prajapati}
\affiliation{National Institute of Standards and Technology, Boulder, Colorado 80305, USA}

\author{Alexandra~B.~Artusio-Glimpse}
\affiliation{National Institute of Standards and Technology, Boulder, Colorado 80305, USA}

\author{Samuel~Berweger}
\affiliation{National Institute of Standards and Technology, Boulder, Colorado 80305, USA}

\author{Matthew~T.~Simons}
\affiliation{National Institute of Standards and Technology, Boulder, Colorado 80305, USA}

\author{Dangka~Shylla}
\affiliation{Associate of the National Institute of Standards and Technology, Boulder, Colorado 80305, USA}
 \affiliation{Department of Physics, University of Colorado, Boulder, Colorado 80309, USA}

 \author{William~J.~Watterson}
\affiliation{Associate of the National Institute of Standards and Technology, Boulder, Colorado 80305, USA}
 \affiliation{Department of Physics, University of Colorado, Boulder, Colorado 80309, USA}

\author{Charles~Patrick}
\affiliation{Associate of the National Institute of Standards and Technology, Boulder, Colorado 80305, USA}
 \affiliation{Department of Physics, University of Colorado, Boulder, Colorado 80309, USA}

\author{Adil~Meraki}
\affiliation{Associate of the National Institute of Standards and Technology, Boulder, Colorado 80305, USA}
 \affiliation{Department of Physics, University of Colorado, Boulder, Colorado 80309, USA}

\author{Rajavardhan Talashila}
\affiliation{Associate of the National Institute of Standards and Technology, Boulder, Colorado 80305, USA}
 \affiliation{Department of Electrical Engineering, University of Colorado, Boulder, Colorado 80309, USA}

\author{Amanda Younes}
\affiliation{Department of Physics and Astronomy, University of California, Los Angeles, California, 90095,
USA}

\author{David S. La Mantia}
\affiliation{National Institute of Standards and Technology, Gaithersburg, MD 20899 USA}

\author{Christopher~L.~Holloway}

\affiliation{National Institute of Standards and Technology, Boulder, Colorado 80305, USA}
\date{\today}

\begin{abstract}
Rydberg states of alkali atoms are highly sensitive to electromagnetic radiation in the GHz-to-THz regime because their transitions have large electric dipole moments. Consequently, environmental blackbody radiation (BBR) can couple Rydberg states together at $\mu$s timescales. Here, we track the BBR-induced transfer of a prepared Rydberg state to its neighbors and use the evolution of these state populations to characterize the BBR field at the relevant wavelengths, primarily at 130 GHz. We use selective field ionization readout of Rydberg states with principal quantum number $n\sim30$ in $^{85}$Rb and substantiate our ionization signal with a theoretical model.  With this detection method, we measure the associated blackbody-radiation-induced time dynamics of these states, reproduce the results with a simple semi-classical population transfer model, and demonstrate that this measurement is temperature sensitive with a statistical sensitivity to the fractional temperature uncertainty of 0.09~Hz$^{-1/2}$, corresponding to 26~K$\cdot$Hz$^{-1/2}$ at room temperature. This represents a calibration-free SI-traceable temperature measurement, for which we calculate a systematic fractional temperature uncertainty of 0.006, corresponding to 2~K at room temperature when used as a primary temperature standard.
\end{abstract}

\maketitle

Radiation thermometry is used widely in science and engineering, including in remote sensing, weather prediction, and manufacturing, among others.
Typical classical radiation thermometers require calibration, mostly using the classic blackbody cavity, which emits a known amount of radiation given its temperature through Planck's law.
This arduous calibration process not only requires calibrating the radiation thermometers, but also the blackbody itself and the contact thermometers used to measure the blackbody's temperature.
As such, the calibration is subject to offsets and errors~\cite{Carter2006}.
However, the redefinition of the SI in 2019 has opened new opportunities for calibration-free radiation thermometers that directly realize the kelvin.
Here, we demonstrate such a quantum-based thermometer based on blackbody radiation-induced transitions within Rydberg atoms.

Rydberg states of alkali atoms have been used as electric field sensors in a comprehensive set of application spaces because their large transition dipole moments make them sensitive to radiation in the GHz-to-THz range \cite{Fan_2015, Sedlacek2012,9748947,NoahNature}.
Likewise, Rydberg atoms are also sensitive to blackbody radiation (BBR) at those same frequencies.
Blackbody radiation has three predominant effects on Rydberg atoms~\cite{PhysRevLett.42.835, PhysRevA.23.2397}.
First, it induces a common AC Stark shift amongst the Rydberg levels, first measured by Hollberg and Hall~\cite{PhysRevLett.53.230}.
This shift, roughly 2.4~kHz at room temperature, has been proposed as a potential thermometer for optical clocks~\cite{PhysRevLett.107.093003}; currently the ambient BBR represents the largest contribution to an optical clock's uncertainty budget \cite{doi:10.1126/science.1153341,PhysRevLett.113.260801} and is quantified using contact thermometers and models of the surrounding environment's emissivity~\cite{PhysRevLett.113.260801}.
Second and third, blackbody radiation induces ionization~\cite{PhysRevA.26.1490, Beterov_2009} and transitions between different Rydberg states~\cite{FIGGER1980,Spencer1982,PhysRevA.51.4010}.
Reference \cite{Norrgard_2021} found that the relative statistical temperature sensitivity $\sigma_T/T $ can approach $1/\sqrt{N_\textrm{Ryd}}$, where $N_\textrm{Ryd}$ is the number of Rydberg atoms, by measuring any of the above three effects.

Recently, we realized a radiation thermometer based on monitoring fluorescence from BBR-populated states in an optically excited vapor cell with a statistical uncertainty as low as $\sigma_T/T =4\times 10^{-4}$~\cite{vaporcellBBR}.
It operated near the peak of the blackbody spectrum by monitoring optical fluorescence induced by BBR excitation at 24.6~THz.
In the microwave regime, we reinterpreted the state population data of Ref.~\cite{PhysRevA.51.4010} taken by state selective field ionization (SFI) to derive the radiometric temperature around the frequency of 167~GHz with an overall statistical uncertainty of about 2~\%~\cite{NoahNature}.

\begin{figure*}[ht]
\raggedright{
a)\hspace{.5cm}\includegraphics[width = .27\textwidth]{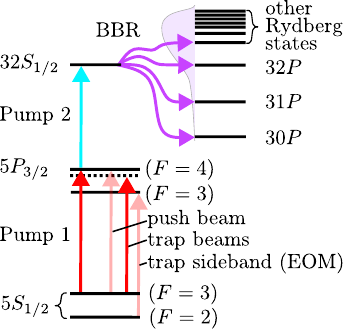} \hspace{1cm} b)\includegraphics[width = .55\textwidth]{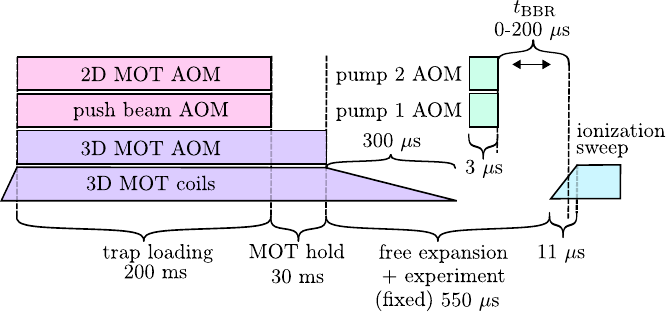}\\[5pt]
c)\hspace{.5cm}\includegraphics[width = .9\textwidth, trim = 0 .05cm 0 0, clip]{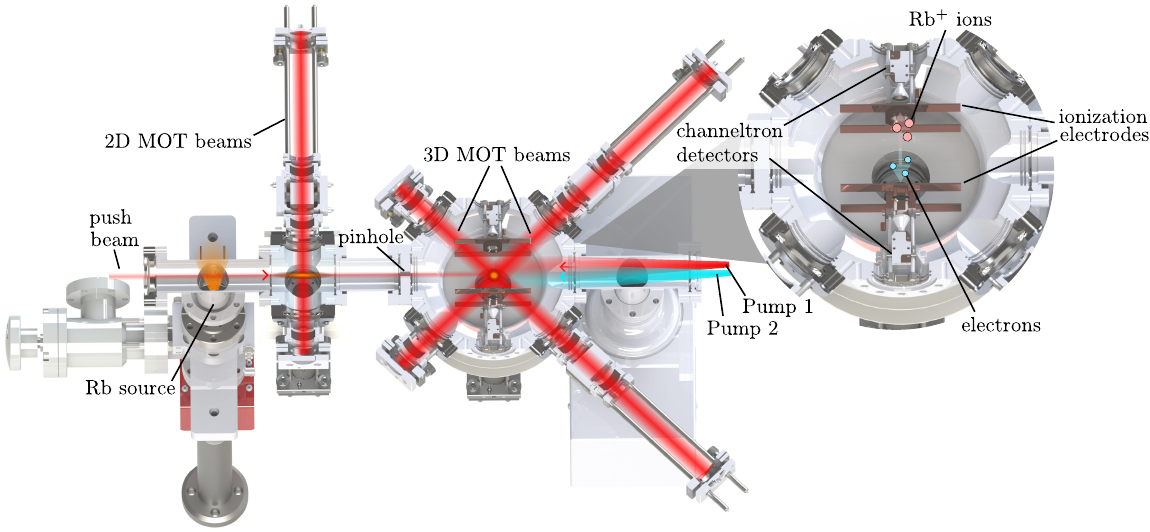}}
\caption{The experimental scheme for measuring BBR-induced state transfer. a) Energy level diagram of the experiment. b) Timing sequence of the experiment. c) Physical layout of the experiment. Not shown is a retro-reflected vertical beam in each of the 2D and 3D MOTs and the anti-Helmholtz coils of the 3D MOT, which are aligned vertically (out of the page).}
\label{fig:ExpLayout}
\end{figure*}

The radiation thermometer described here  is conceptually similar to the proof-of-principle demonstrated in Refs.~\cite{PhysRevA.51.4010,NoahNature}, while being optimized for radiometric thermometry and using cold atoms to suppress collisional systematics. 
We realize the kelvin by tracking the BBR-induced transfer from the $32S$ state of $^{85}$Rb to its neighboring states, particularly to the $32P$ state.
This transition is sensitive to BBR radiation at a frequency of 130~GHz. 
Such microwave frequencies have posed a unique challenge in radiometry, as the BBR energy density is roughly 2000 times lower than the peak around 30~THz at 300~K.
The transition rate between $32S$ and $32P$ is essentially the product of the BBR energy density, the transition dipole moments, and SI constants, meaning they represent an SI-traceable primary temperature measurement \cite{Norrgard_2021}.

The measurement scheme relies on several well-developed techniques and is as follows: We prepare a sample of roughly $10^6$ $^{85}$Rb atoms at roughly 1~mK in a magneto-optical trap (MOT).
We then pulse a two-photon excitation to a Rydberg state.
We wait a time $t_\textrm{BBR}<100$~$\mu$s for blackbody radiation to couple from this Rydberg state to other states~\cite{PhysRevA.51.4010, PhysRevA.105.063104}.
Next, we sweep an electric field to selectively ionize Rydberg state atoms~\cite{gallagher1994rydberg} and collect the ions and stripped electrons using electron avalanche detectors.
Each measurement takes a time $\tau_{\rm shot}$ of 354~ms, consisting of 231~ms of experiment (Fig. \ref{fig:ExpLayout}b) and 123~ms of dead time.

Figure~\ref{fig:ExpLayout} details the relevant energy level diagram (a), the timing (b), and the physical layout (c) of the experiment.
We pass 4~A of current through a resistive dispenser filled with solid Bi$_{2}$Rb$_{3}$ alloy, which then heats to its vaporization threshold and releases pure Rb.
The Rb vapor makes its way across the vacuum chamber into the center of a 2D MOT, consisting of two retro-reflected laser beams ($-20$~MHz detuned from the D$_2$ line ($5S_{1/2} \rightarrow 5P_{3/2}$) with $\sim$100~mW of power in a 1~cm one-$\sigma$ beam radius) and four permanent bar magnets.
The atoms are collected here, and a weak push beam (resonant to the D$_2$ line with $\sim$100~$\mu$W power in a 2~mm one-$\sigma$ beam radius) pushes the atoms through a 4~mm pinhole into a 3D MOT.
Loading from the 2D to the 3D MOT through a pinhole allows for the 3D MOT chamber to be free of background room-temperature atoms.
The 2D MOT chamber is evacuated below $1.3 \times 10^{-9}$ Pa, and the 3D MOT chamber is evacuated to $7.1 \times 10^{-8}$ Pa.

The 3D MOT consists of three retro-reflected laser beams ($-20$~MHz detuned from the D$_2$ line with $\sim$80~mW of power in a 1~cm one-$\sigma$ beam radius) and two coils in an anti-Helmholtz configuration.
The current through the coils is controlled with an insulated-gate bipolar transistor which allows the field to be switched off in $\sim$ 300~$\mu$s.
We estimate the cloud to contain $\sim$$2\times 10^{6}$ atoms, of which $\sim$5400 participate in the measurement, with a temperature of $\sim$ 0.5~mK (details in supplemental material). 

After the trap is released, the atoms are excited to a Rydberg state via Pump 1 (resonant to the D$_2$ line with $\sim9$~mW in a 5~mm one-$\sigma$ beam  radius) and Pump 2 (resonant on the $5P_{3/2}\rightarrow 32S_{1/2}$ transition with 57~mW in a 5~mm one-$\sigma$ beam radius, locked to a two-photon electromagnetically induced transparency in a reference cell).
After the blackbody  coupling time $t_\textrm{BBR}$, ionization is performed with two electrodes placed 56~mm apart that are swept from 0~kV to 3~kV in $\sim 7~\mu$s,  and the ions and their electrons are collected using channel electron multiplier (CEM), or ``channeltron'', detectors.

The current incident on the anode of the CEM is converted into a voltage using a transimpedance amplifier with a gain of $10^3$~V/A and recorded on an oscilloscope. The circuit driving the ionization electrodes is diagrammed in the supplemental material. The initial state of $32S_{1/2}$ is chosen for experimental reasons. Generally, lower $n$ states are more resolvable with SFI, but require a larger electric field to ionize. An initial Rydberg state of $32S_{1/2}$ is chosen because it is the lowest state that ionizes optimally given the timing constraints of the experiment. 

\begin{figure}
a)\hspace{.00\linewidth}\includegraphics[width = .94\linewidth, trim = .25cm .15cm .2cm .26cm, clip]{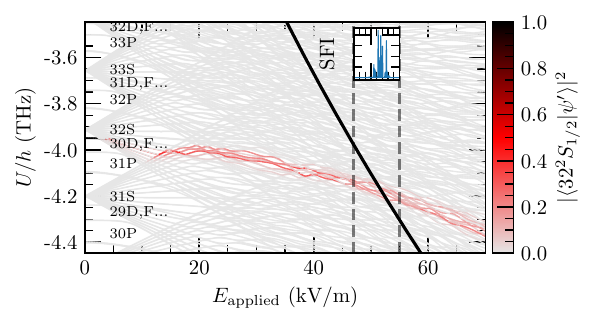}\\[1pt]
b)\includegraphics[width = .95\linewidth, trim = 0 .25cm 0 .3cm]{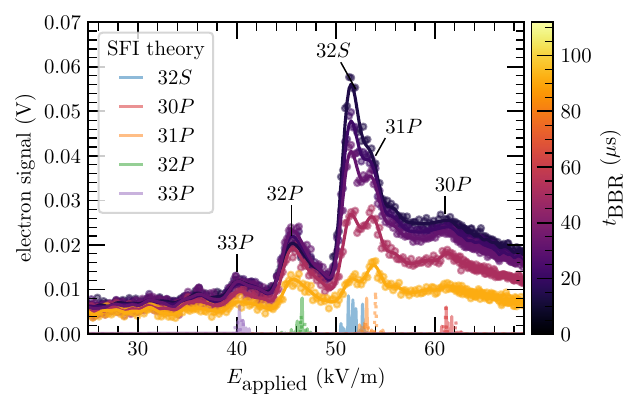}
\caption{Theoretical model of SFI. a) The Stark map is calculated for a ramp from 0~kV to 70~kV/m in $11~\mu$s, and for each point in the ramp the overlap of each state with the original state $32S_{1/2}$ is shown in the colormap.  The ionization energy is drawn as a solid black line, and the resulting expected SFI signal is shown in an inset. b) An equivalent calculation is performed for each relevant state, and the measured electron ionization signal at various $t_\textrm{BBR}$ is overlayed. For the $P$ state theory, solid lines indicate $P_{3/2}$ state contributions while dotted lines represent $P_{1/2}$ state contributions.}
\label{fig:SFI}
\end{figure}

To make sense of the resulting SFI signal, we use a theoretical model~\cite{gallagher1994rydberg, PhysRevA.98.063404}.
We produce a Stark map of all relevant states as the electric field is swept and propagate the projection of the initial state onto the eigenstates at each time step.
At each point in the sweep, a Stark eigenstate will be ionized if its classical ionization field $E_\text{ionization}$ is reached, calculated as \cite{PhysRevA.17.1226}
\begin{equation}
E_\text{ionization} = \frac{m_e a_0}{4 e \hbar^2}U_{i}^2,
\end{equation}
where $U_{i}$ is the energy of the state below the free electron continuum, $m_e$ is the mass of the electron, $a_0$ is the Bohr radius, and $\hbar$ is the reduced Planck constant. Each Stark eigenstate will contribute to the SFI signal when its ionization energy is reached, with a relative amplitude corresponding to the likelihood of arriving in that Stark eigenstate given the initial state. Such a calculation for $32S_{1/2}$ is shown in Fig.~\ref{fig:SFI}a.

Both the ions and the electrons produced by the electric field ramp are collected. The ion signal is considerably stronger than the electron signal, and the electron signal requires capacitive readout which limits the detection bandwidth.  However, the electrons arrive near-instantaneously, while the ions' mass places their transit time at timescales of order the length of the ionization ramp. Therefore, we use the electron signal with a well defined applied field to time relationship to identify peaks, and the ion signal to count state populations. 

The field at which ionization occurrs is calculated from the electron time-of-arrival by recording the voltages applied to each electrode with a high voltage probe and using a finite element simulation of the experiment to convert voltages at the electrodes to fields at the center of the trap (details in supplemental material). The electron signal trace over a span of 100 $\mu$s of BBR-induced coupling after populating $32S$ is shown in Fig.~\ref{fig:SFI}b. Overlayed are the theoretical SFI contributions from the relevant states, each calculated as in Fig.~\ref{fig:SFI}a. Agreement between the theoretical fields at which ionization occur and the measured ionization signal demonstrates adequate understanding of the field ionization process and confidently identifies each detected peak.

\begin{figure}
a)\hspace{.02\linewidth}\includegraphics[width = .92\linewidth, trim = .1cm .2cm .1cm .25cm, clip]{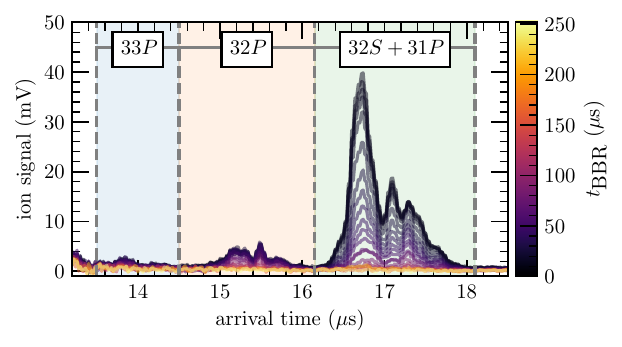}\\
b)\includegraphics[width = .87\linewidth,trim = .1cm .2cm 0 .2cm, clip]{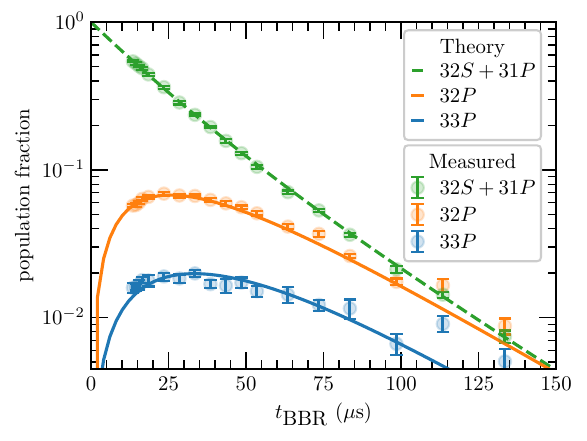}\hspace{.08\linewidth}
\caption{Population tracking of the ionized states as a function of the BBR evolution time $t_\text{BBR}$ in a room-temperature (296~K) environment. a) Time dynamics of the ion signal, with each peak labeled with the corresponding quantum state. b) Each peak is integrated to count the atom population in each state. Each data point represents 40 measurements. The solid lines represent theoretical time dynamics from Eq. \ref{eq:timedynamicsol}.}
\label{fig:timedynamicssol}
\end{figure}

Population in each state can be counted by integrating each peak on the ion signal, which has sufficient bandwidth to fully separate each peak (Fig. \ref{fig:timedynamicssol}a). The saturation and kinetic energy-dependent gain of the detector are characterized in the supplemental material and corrections are applied before integrating the signal to count ions. The ion signal at various blackbody evolution times $t_\text{BBR}$ is shown in Fig.~\ref{fig:timedynamicssol}b. Initially, the $32S$ state is populated by the pump lasers.  Blackbody radiation then couples population from $32S$ to nearby Rydberg states that are dipole-allowed transitions, and eventually all population decays to states not accessible via SFI readout.

A theoretical model for the BBR-induced state transfer can be constructed analytically~\cite{PhysRevA.51.4010}. Because natural decay and blackbody-driven couplings are both incoherent processes, we can model the time dynamics of the system using a semi-classical rate equation model. If the transfer rate from a state $\ket{i}$ to a state $\ket{j}$ induced by blackbody radiation and natural decay is given by $\Gamma_{i\rightarrow j}$ and the lifetime of a state $\ket{i}$ from all such couplings is $\tau_{i}$, then for a set of $N$ states with populations $n_{i}$ the time dynamics are given by
\begin{equation}
    \frac{\partial}{\partial t} \begin{bmatrix} n_1 \\ n_2 \\ ... \\ n_N \end{bmatrix} = \underbrace{
    \begin{bmatrix}
    -1/\tau_1 & \Gamma_{2\rightarrow 1} & ... & \Gamma_{N\rightarrow 1} \\
    \Gamma_{1\rightarrow 2} & -1/\tau_2 & ... & \Gamma_{N\rightarrow 2} \\
    ... & ... & ... & ... \\
    \Gamma_{1 \rightarrow N} & \Gamma_{2 \rightarrow N} & ... & -1/\tau_N
    \end{bmatrix}}_{\equiv M} \begin{bmatrix} n_1 \\ n_2 \\ ... \\ n_N  \end{bmatrix}\label{eq:matrix},
\end{equation}
where the off-diagonal matrix elements represent individual couplings and the diagonal matrix elements represent all transfer out of a state. This differential equation has the solution
\begin{equation}
     \begin{bmatrix} n_1 (t)\\ n_2 (t)\\ ... \\ n_N (t) \end{bmatrix} = e^{M t}  \begin{bmatrix} n_1 (t=0)\\ n_2 (t=0)\\ ... \\ n_N (t=0)\end{bmatrix}.
     \label{eq:timedynamicsol}
\end{equation}
States with principal quantum numbers from $n$=5 through $n$=50 and angular momentum quantum numbers $\ell$=0 through $\ell$=20 are considered, with the initial state taken to be that with only $\ket{32S}$ populated. The BBR and decay-induced transition rates $\Gamma_{i\rightarrow j}$ and lifetimes $\tau_i$ contained in the matrix $M$ are calculated using the ARC python package \cite{ARC}.
Assuming that only the $\ket{32S_{1/2}}$ state is initially populated, the solution at room temperature is compared with measurement in Fig.~\ref{fig:timedynamicssol}b. Note that Rydberg $P_{1/2}$ and $P_{3/2}$ states are included separately in the model, and the ``$P$ state'' populations are taken to be the sum of these two. Agreement with the theoretical time dynamics predicted by Eq. \ref{eq:timedynamicsol} is achieved with an overall normalization as well as allowing an offset in $t_\textrm{BBR}$ due to finite optical pumping and ionization ramp times.

To turn state population dynamics into a primary BBR temperature measurement, it is necessary to make a direct comparison of measured values to theory, without the need for scaling factors. One way to parameterize the time dynamics in a dimensionless way is with the ratio $\mathcal{R}$ of the populations in the two measured peaks:
\begin{equation}
    \mathcal{R} \equiv \frac{N_{\ket{32P}}}{N_{\ket{32S} }+N_{\ket{31P}}}.
\end{equation}
The value of $\mathcal{R}$ at any time is approximately linearly proportional to the environmental temperature, as at higher temperatures there is a higher BBR photon density to couple $32S$ to $32P$. This ratio is a temperature-sensitive metric that is independent of fluctuations in the trap loading and in the preparation efficiency of the initial Rydberg state.

To demonstrate the temperature sensitivity of this measurement, the vacuum chamber surrounding the 3D MOT was wrapped in multilayer aluminum foil and heated with a resistive element. The heaters are held at constant power for at least eight hours before each measurement to establish thermal equilibrium. The temperature of the environment is measured with resistance temperature detectors at four locations on the surface of the vacuum chamber, and the environmental temperature $T_\textrm{env}$ is taken to be the mean of these measurements. Taking the temperature at the surface of the chamber is justified because the only heat loss from the inside of the chamber is via radiation through a small solid angle provided by the viewports. 

\begin{figure}
\includegraphics[width = \linewidth ,trim = 0 0cm 0 .25cm, clip]{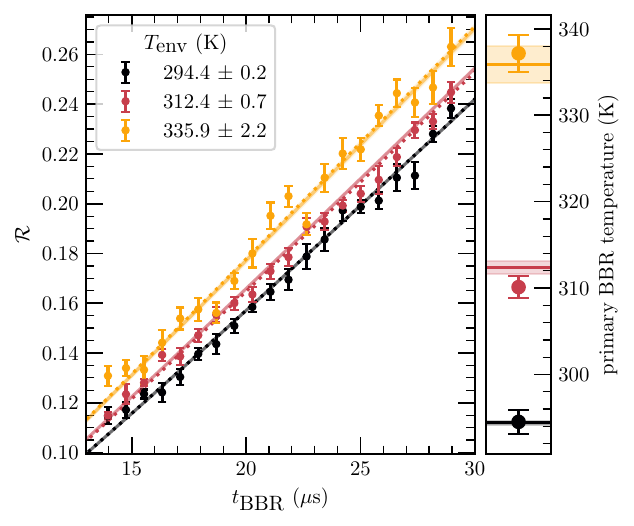}
\begin{tabular}{c|c|c}
\hspace*{.2cm}RTD temps. (K) \hspace*{.2cm}& \hspace*{.3cm}$T_\textrm{env}$ (K)\hspace*{.3cm} & \hspace*{.2cm}BBR temp. (K)\hspace*{.2cm}\\

\hline
294, 294, 295, 294 &294.4 $\pm$ 0.2 & 294.5 $\pm$ 1.4\\
312, 313, 314, 310 & 312.4 $\pm$ 0.7 & 310.1 $\pm$ 1.3\\
337, 336, 341, 329  & 335.9 $\pm$ 2.2 & 337.2 $\pm$ 2.1 

\end{tabular}
\caption{Demonstration of primary thermometry as the vacuum chamber is heated to various temperatures. Left: The peak ratio $\mathcal{R}$ as a function of blackbody evolution time $t_\textrm{BBR}$. Points with error bars are measured values with each point representing 40 measurements.  Solid lines indicate theoretical predictions using Eq. \ref{eq:timedynamicsol} and the classically measured temperature $T_\textrm{env}$, with a shaded region representing the uncertainty of the classical measurement. Dotted lines represent temperature-floated fits based on Eq. \ref{eq:timedynamicsol} that are used to determine the primary BBR temperature measurement. Right: The result of the temperature fits. Each primary temperature measurement is represented as a point with errorbars associated to the statistical uncertainty of the fit. The solid lines represent the associated classically measured $T_\textrm{env}$.}
\label{fig:tempdependence}
\end{figure}

The resulting population ratios measured are shown in Fig.~\ref{fig:tempdependence}.
Each trace of $\mathcal{R}$ is turned into a primary temperature measurement by  fitting it to a theoretical value of $\mathcal{R}(t_\textrm{BBR}, T)$ calculated from Eq. \ref{eq:timedynamicsol}, with the temperature floated.

The measurement is most sensitive in the first $\sim20~\mu$s of BBR evolution time because all relavent states are most populated in this period. Each trace in Fig. \ref{fig:tempdependence} represents a 4.7 minute measurement which yields a mean of 1.6~K of statistical temperature uncertainty. 

A summary of systematic and statistical uncertainties are provided in Table \ref{tab:uncertaintybudget}, where $\sigma_T/T$ is the fractional uncertainty of the measured temperature in Kelvin, and $\mathcal{S}(\sigma_T/T) \equiv \sigma_T/T \cdot \sqrt{\tau_\textrm{meas}}$ is the statistical sensitivity. Here, $\tau_\textrm{meas}$ is the total measurement time including dead time. Detailed calculations for each systematic as well as the statistical sensitivity are included in the supplemental material. We find that as a primary (calibration-free) temperature measurement, our system has a fractional systematic uncertainty of 0.006 representing 2~K of absolute temperature uncertainty at room temperature.

\begin{table}[h]
    \begin{tabular}{|p{5cm} |c|}
    \hline
        \textbf{systematic uncertainty} & $\sigma_T/T$\\ \hline
        detector non-linearity & 0.002\\
        ion time-of-flight overlap & 0.005 \\
        determination of $t_\textrm{BBR} = 0$ & 0.002\\
        detection signal  artefacts & 0.003\\
        time-dependent gain calibration & $1\times 10^{-5}$\\
        \hline
        total (quadrature sum) & 0.006\\
        \hline\hline
        \textbf{statistical senstivity} & $\mathcal{S}(\sigma_T/T)$ ~($ \textrm{Hz}^{-1/2}$) \\ \hline

        total & 0.086 $\pm$ 0.009
        \\
        \hline
    \end{tabular}

    \caption{Uncertainty budget for the temperature measurement.}
    \label{tab:uncertaintybudget}
\end{table}

Current leading atomic clock BBR characterization relies on conventional thermal sensors, usually platinum resistance temperature detectors (RTDs) placed in the environment surrounding the atoms, operating with temperature uncertainties on the order 10~mK giving rise to a fractional frequency uncertainty on the order 10$^{-18}$ for Yb due to the Stark shift induced by the BBR \cite{Heo_2022, PhysRevLett.113.260801}. While the method described in this paper is a more direct temperature measurement as it measures the BBR experienced by the atoms in-situ at their trap location, the associated absolute temperature uncertainty is two orders of magnitude higher than the state of the art. However, this proof-of-principal demonstration can be improved with careful characterization and mitigation of the systematic sources of error presented.

We have described a method for measuring blackbody radiation via tracking the population of Rydberg states in cold atoms. We have demonstrated selective field ionization to read out $n$$\sim$30 states of $^{85}$Rb and correlated these measurements to theoretical ionization fields. We used this readout scheme to measure BBR-induced state transfer and decay in Rydberg states, providing a semi-classical model for the time dynamics, and demonstrated that this measurement is temperature-sensitive. We resolve 1.6~K in 4.7 minutes, with an absolute uncertainty of 2~K. This work represents a proof-of-concept for using Rydberg atoms as an SI-traceable radiation thermometer.

\vspace{.3cm}
\begin{acknowledgments}
\textit{Acknowledgements}---This work was funded by the National Institute of Standards and Technology (NIST) through the NIST-on-a-Chip (NOAC) and through the Innovations in Measurement Science (IMS) program.  

\textit{Conflicts of interest}---The authors have no conflicts to disclose.

\textit{Data availability}---The data relevant to the findings of this work is
available at \href{https://doi.org/10.18434/mds2-3499}{doi:10.18434/mds2-3499}.
\end{acknowledgments}
\nocite{suppMats}
\nocite{CloudTempMeas}
\nocite{Rb85D2Lifetime}
\nocite{PhysRevLett.93.063001}
\nocite{Norrgard_2021}

\end{document}


\preprint{APS/123-QED}

\title{Primary quantum thermometry of mm-wave blackbody radiation  via \\induced state transfer in Rydberg states of cold atoms}

\author{Noah~Schlossberger}
\email{noah.schlossberger@nist.gov}
\affiliation{National Institute of Standards and Technology, Boulder, Colorado 80305, USA}

\author{Andrew~P.~Rotunno}
\affiliation{National Institute of Standards and Technology, Boulder, Colorado 80305, USA}

\author{Stephen~P.~Eckel}
\affiliation{National Institute of Standards and Technology, Gaithersburg, MD 20899 USA}

\author{Eric~B.~Norrgard}
\affiliation{National Institute of Standards and Technology, Gaithersburg, MD 20899 USA}

\author{Dixith~Manchaiah}
\affiliation{Associate of the National Institute of Standards and Technology, Boulder, Colorado 80305, USA}
\affiliation{Department of Physics, University of Colorado, Boulder, Colorado 80309, USA}

\author{Nikunjkumar~Prajapati}
\affiliation{National Institute of Standards and Technology, Boulder, Colorado 80305, USA}

\author{Alexandra~B.~Artusio-Glimpse}
\affiliation{National Institute of Standards and Technology, Boulder, Colorado 80305, USA}

\author{Samuel~Berweger}
\affiliation{National Institute of Standards and Technology, Boulder, Colorado 80305, USA}

\author{Matthew~T.~Simons}
\affiliation{National Institute of Standards and Technology, Boulder, Colorado 80305, USA}

\author{Dangka~Shylla}
\affiliation{Associate of the National Institute of Standards and Technology, Boulder, Colorado 80305, USA}
 \affiliation{Department of Physics, University of Colorado, Boulder, Colorado 80309, USA}

 \author{William~J.~Watterson}
\affiliation{Associate of the National Institute of Standards and Technology, Boulder, Colorado 80305, USA}
 \affiliation{Department of Physics, University of Colorado, Boulder, Colorado 80309, USA}

\author{Charles~Patrick}
\affiliation{Associate of the National Institute of Standards and Technology, Boulder, Colorado 80305, USA}
 \affiliation{Department of Physics, University of Colorado, Boulder, Colorado 80309, USA}

\author{Adil~Meraki}
\affiliation{Associate of the National Institute of Standards and Technology, Boulder, Colorado 80305, USA}
 \affiliation{Department of Physics, University of Colorado, Boulder, Colorado 80309, USA}

\author{Rajavardhan Talashila}
\affiliation{Associate of the National Institute of Standards and Technology, Boulder, Colorado 80305, USA}
 \affiliation{Department of Electrical Engineering, University of Colorado, Boulder, Colorado 80309, USA}

\author{Amanda Younes}
\affiliation{Department of Physics and Astronomy, University of California, Los Angeles, California, 90095,
USA}

\author{David S. La Mantia}
\affiliation{National Institute of Standards and Technology, Gaithersburg, MD 20899 USA}

\author{Christopher~L.~Holloway}

\affiliation{National Institute of Standards and Technology, Boulder, Colorado 80305, USA}
\date{\today}

\date{\today}
\maketitle
\onecolumngrid
\section*{Supplemental material}

\subsection{Atom trap characteristics} \label{subsec:tempmeas}
The temperature of the cloud can be deduced from a free expansion of the cloud.
The time evolution of the Gaussian width $\sigma$ of the cloud is given by \cite{CloudTempMeas}
\begin{equation}
    \sigma^2(t) = \sigma^2(t=0) + \frac{k_B T}{m}t^2, \label{eq:expansionRate}
\end{equation}
where $k_B$ is the Boltzmann constant, $m$ is the mass of the Rb atoms, and $T$ is the temperature of the atoms.

\begin{figure}[h]
\includegraphics[width = .8\linewidth]{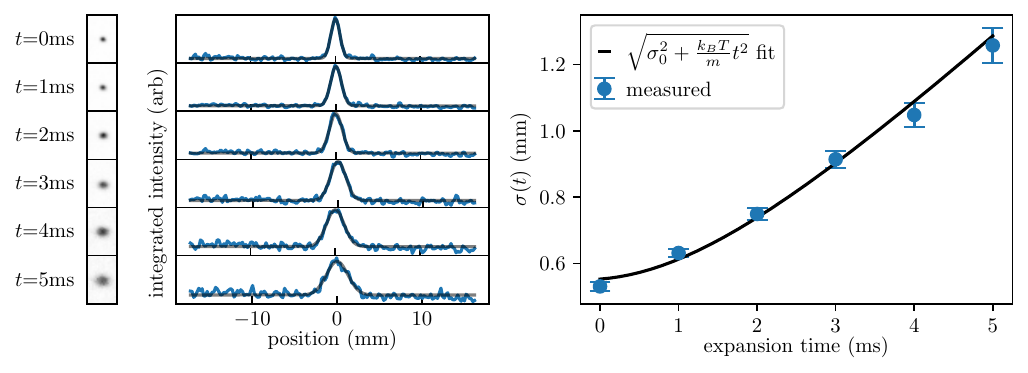}
\caption{Free expansion of the atom cloud after the 3D MOT is turned off. Top left: images of the atoms at various expansion times. Top right: the images are integrated along one axis to give the horizontal Gaussian profile (blue), which is then fit (black). Bottom: The width from each Gaussian fit form an expansion profile which is fit to Eq. \ref{eq:expansionRate}.}
\label{fig:freeExpand}
\end{figure}

The temperature of the atoms is measured by releasing the 3D MOT and allowing the atoms to freely expand. After the free expansion time, the atoms are illuminated with resonant probe light and the fluorescence is imaged by a focused complementary metal-oxide-semiconductor (CMOS) camera 30~cm away from the atoms. The camera is exposed for 100~$\mu$s, chosen to be fast compared to expansion time but long enough to allow for adequate signal-to-noise. The pixels on the image are converted to spatial size in the atom's plane by projecting the geometry of vacuum features in the same image of known size. The result of a free-expansion measurement is shown in Fig.~\ref{fig:freeExpand}.

The fit yields an initial width of $0.56\pm0.01$~mm and a temperature of $0.54\pm0.03$~mK.

The atom number is determined by imaging the fluorescence of the cloud in the last 20~ms before the trap is released. Because the Rabi rate of the trapping light is large compared to the natural decay rate $\Gamma = 2\pi \times 6$~MHz\cite{Rb85D2Lifetime} of the D2 transition, each atom will emit approximately $\frac{\Gamma}{2\pi}$ photons per second isotropically. The cloud is imaged using a charge-coupled device (CCD) camera a distance $d_\text{CCD}$ of 30~cm from the atoms with an effective aperture radius $a_\text{CCD}$ of 2~cm. The power-to-count ratio of the CCD camera at a wavelength of $\lambda = 780$~nm is calibrated by directly imaging a laser beam of known power. The power reaching the detector is given by
\begin{equation}
    P_\text{CCD} \approx N_\text{atoms}\underbrace{\left(\frac{\Gamma}{2\pi} \frac{h c}{\lambda}\right)}_\text{power per atom} \underbrace{\left(\frac{\pi a_\text{CCD}^2}{4\pi d_\text{CCD}^2}\right)}_\text{collection efficiency}.
\end{equation}
From the measured power on the CCD detector we can extract the atom number, yielding a typical number of $N_\text{atoms}$ = $2\times 10^6$.

When populating the $32S_{1/2}$ state with the pump lasers, the number of excited atoms could be limited by the Rydberg blockade effect. If the density of excited atoms becomes too high, the van der Waals interaction between nearby Rydberg states will give rise to a frequency shift $\Delta$ given by \cite{PhysRevLett.93.063001}
\begin{equation}
    \Delta = -C_6 / r^6,
\end{equation}
where $C_6$ is the Rydberg-Rydberg van der Waals coefficient and $r$ is the distance between the atoms. If the magnitude of this shift is larger than the linewidth of the excitation (in the power-broadened regime, this is approximately equal to the excitation Rabi rate $\Omega$), than this will prevent further excitation into Rydberg states. The Rydberg blockade radius $r_b$, and therefore the maximum Rydberg density before a blockade occurs $n_b$, are given by
\begin{equation}
\begin{cases}
    r_b= \sqrt[\leftroot{-2}\uproot{2}6]{\frac{|C_6|}{\Omega_\text{pump 2}}} \\
    n_b = \frac{1}{\frac{4}{3}\pi r_b^3}
    \end{cases},
\end{equation}
where $\Omega_\text{pump 2}$ is the Rabi rate of the photon coupling into the Rydberg state (where we have ignored the first photon because $\Omega_\text{pump 1}$ is large compared to $\Omega_\text{pump 2}$). For $32S_{1/2}$, the van der Waals parameter $C_6$ is equal to $-0.040$~GHz~$\mu$m$^6$, calculated perturbatively using ARC. If the power of the Pump 2 laser is 57~mW in a 5~mm one-$\sigma$ beam width, then $\Omega_\text{pump 2} = 2\pi \times 194$~kHz.

The maximum number is set by the size of the cloud. The atomic density is given by
\begin{align}
    n(\vec r) &= N_0 \frac{1}{(2\pi)^{3/2} \sigma_x \sigma_y \sigma_z}e^{-\frac{x^2}{2\sigma_x^2}-\frac{y^2}{2\sigma_y^2}-\frac{z^2}{2\sigma_z^2}}\\
    & \approx N_0 \frac{1}{(2\pi)^{3/2}\sigma^3}e^{-\frac{|r|^2}{2\sigma^2}},
\end{align}
where $\sigma$ is the width of the cloud, and we have taken the cloud to be spherically symmetric. We can then make the assumption that the Rydberg atom density will have a similar distribution after pumping.
Then in order to not reach the Rydberg blockade, the maximum total number of Rydberg atoms $N_r$ is given by
\begin{equation}
    N_r = (2\pi)^{3/2}\sigma^3 n_b.
\end{equation}
Given the measured cloud size $\sigma$ of 0.56$\pm$.01~mm, the maximum number of atoms in the Rydberg state in the cloud is $\sim 5\times 10^7$. As the measured number of ground-state atoms is lower than this and the two-photon excitation has a limited efficiency, we can assume that a blockade will not occur.

To determine the number of atoms in the Rydberg state, we can refer the gain of the detector at -1450V to the gain of the detector at -2500V in an unsaturated low-atom regime and use the manufacturer specifications of the channeltron detector to infer the gain. The integral of the ion signal can be estimated as
\begin{equation}
     \int V(t)dt = \underbrace{(G_\textrm{detector} \varepsilon_\textrm{detector} N_\textrm{ions} e)}_\textrm{charge deposited on anode} Z_\textrm{amp},
\end{equation}
where $G_\textrm{detector}$= 5$\times 10^4$ is the gain 
 of the electron avalanche at -1450V (referenced to the specified value of $4\times 10^7$ at -2500V), $\varepsilon_\textrm{detector}$= 0.4 is the velocity-dependent efficiency of the detection , $e$ is the charge of an electron, and  $Z_\textrm{amp}$= $10^3$ V/A is the gain of the trans-impedance amplifier . The number of ions is then
 \begin{equation}
     N_\textrm{ions} = (312 \;\textrm{mV}^{-1}\cdot\mu\textrm{s}^{-1}) \left(\int V(t)dt\right). 
 \end{equation}
 At the earliest measured times of $t_\textrm{BBR}=12$~$\mu$s, we typically measure a total signal of 34~mV$\cdot\mu$s, corresponding to approximately 5400 ions.
\subsection{Selective field ionization}
\begin{figure}[h]
a)\includegraphics[width = .31\linewidth, trim= 30 0 0 -30,clip, valign = t]{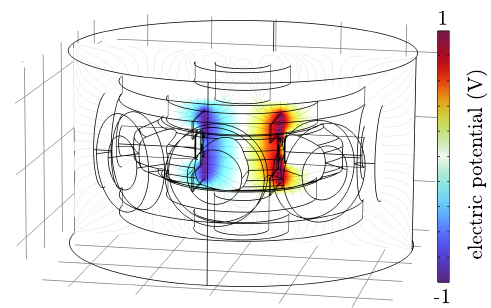}b)\includegraphics[width = .44\linewidth, valign = t]{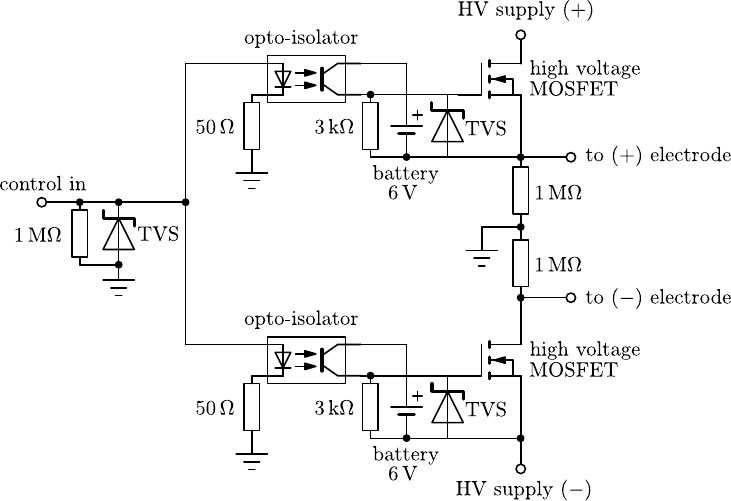} c)\includegraphics[width = .195\linewidth, valign = t]{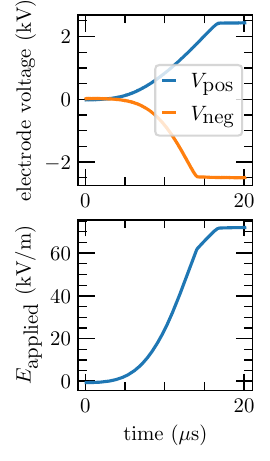}
\caption{a) A COMSOL simulation of the electric potential in the atom trap due to the ionization electrodes. $\pm$ 1V is applied to each electrode. b) Circuit to apply high voltage pulses to the ionization electrodes based on MOSFET switching. c) Top: measured voltages on the electrodes during a sweep. The high voltage supplies are held at $+2500$~V and $-2500$~V. Bottom: the resulting field at the atom location using Eq. \ref{eq:Eapplied}.}
\label{fig:HVcircuit}
\end{figure}

A finite-element model was programmed in COMSOL multiphysics\textsuperscript{\textregistered}\footnote{Certain commercial equipment, instruments, software, or materials, commercial or non-commercial, are identified in
this paper in order to specify the experimental procedure adequately. Such identification does not imply
recommendation or endorsement of any product or service by NIST, nor does it imply that the materials or equipment
identified are necessarily the best available for the purpose.} in order to simulate the electric potential inside of the vacuum chamber. Exact models of the electrodes were used and a simplified model of the octagonal vacuum chamber were used (for example, the threads near each viewport are removed) in order to reduce the dimensionality of the simulation. The solution is shown in Fig. \ref{fig:HVcircuit}a.

The symmetry of the electrode geometry means that the electric field at the center of the trap is directly proportional to the $\textit{difference}$ of the electrode voltages, so an assymetric set of electrode voltages still produces a linear response. The COMSOL simulation indicates that the electric field at the center of the trap is given by 
\begin{equation}
    E_\text{applied} = (14.425\, \text{m}^{-1}) (V_\text{pos} - V_\text{neg}), \label{eq:Eapplied}
\end{equation}
where $V_\text{pos} $ and $ V_\text{neg}$ are the voltages applied to each electrode.

\begin{figure}
    \includegraphics[width = .85\textwidth]{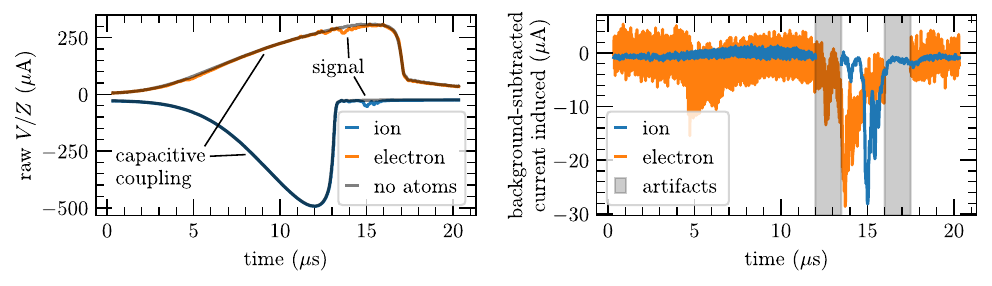}
    \caption{Processing of the raw detection signal. Left: The raw voltage $V$ measured on the channeltron anode divided by the measurement impedance $Z$ (a 50 $\Omega$ resistor for the electron signal and a $10^3$ V/A transimpedance amplifier for the ion signal). The ionization electrode produces a large parasitic capactively coupled background signal. Right: This background signal is subtracted away. Imperfect cancelations of the high-slope regions of the background signal leads to artifacts which are removed, the regions indicated with gray shading.}
    \label{fig:backsubtract}
\end{figure}

The field ionization ramps are generated using high voltage MOSFETs operating as switches between high voltage supplies and a pull-down resistor to ground. The circuit is shown in Fig.~\ref{fig:HVcircuit}b.

The resulting switching times are around 11~$\mu$s, designed to be short compared to the BBR dynamics but long enough to allow the SFI peaks to be resolved given the parasitic capacitance on the CEM detector readout. A pair of high-bandwidth, high-voltage scope probes are used to measure the voltages on the electrodes, such that the electric field seen by the atoms as determined in Eq. \ref{eq:Eapplied} can be found as a function of time. These are shown in Fig.~\ref{fig:HVcircuit}c.

Due to parasitic capacitance between the leads of the ionization plates and the channeltron, the ramping of the ionization plates leads to pickup on the channeltron readout, as shown in Fig. \ref{fig:backsubtract}a. Because this effect is repeatable, we can measure this induced pickup without loading the trap and subtract it to isolate the ion and electron signals, as in Fig. \ref{fig:backsubtract}b. The ion signal is preferable for population counting because the ions arrive after the coupling artifact. The temperature measurements are taken with an asymmetric set of ionization voltages: $V_\textrm{pos}$~=~3200~V and $V_\textrm{neg}$~=~-1300~V. This further increases the ions' time of flight, temporally isolating them from the background signal, which would otherwise contribute substantial systematic error.

\subsection{Channeltron gain calibration}
\begin{figure}[h]
a)\includegraphics[width = .4\textwidth]{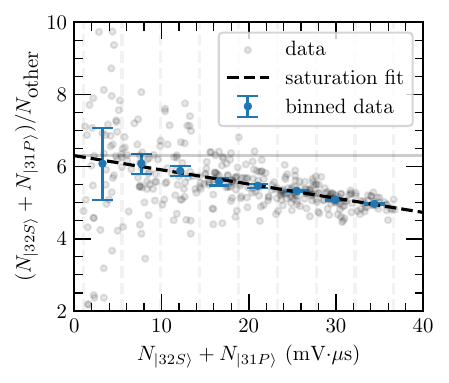}
b)\raisebox{2.7\height}{\begin{tabular}{l l l}
$V_\textrm{det}$(V) & relative gain & $N_\textrm{typ}$/$N_\textrm{sat}$\\
\hline
-1300 & 1 & 0.04\\
-1400 & 3.1 $\pm$ 0.3& 0.05\\
-1500 & 13 $\pm$ 2& 0.15\\
-1600 & 26 $\pm$ 3& 0.22\\
\end{tabular}}
\caption{Saturation of the channeltron detector. a) The saturation is measured with the front plate of the detector held at -1450V. The total atom number is varied by changing the MOT load time. The data is binned and fit to the theoretical form given in Eq. \ref{eq:satfit}. b) A similar measurement is performed at various detector voltages $V_\textrm{det}$. The relative gains are calculated by taking the unsaturated integral of the ($32S + 31P$) peak under identical experimental conditions, and the typical value $N_\textrm{typ}$ of this integral under the conditions used for the primary BBR measurement are compared to the measured $N_\textrm{sat}$ at each voltage.}
\label{fig:sat}
\end{figure}

The channeltron electron avalanche detectors can behave non-linearly in several ways that affect the counting of ions. Firstly, the detector can saturate if the incident flux of ions is too high. The detector is operated with its front plate held at a voltage of $V_\textrm{det}=$-1450~V, significantly below the manufacturer-recommended voltage of -2500~V, in order to minimize saturation.  The saturation value of the peak integral can be estimated by comparing the peak heights as a function of number of ions, assuming that the large peak containing $32S+32P$ is saturating much more than the other peaks which are of order ten times smaller. The saturation at the chosen voltage of -1450~V is shown in Fig. \ref{fig:sat}a.  

Assuming the standard saturation curve:
\begin{equation}
N_\textrm{meas} = \frac{N_\textrm{true}}{1 + N_\textrm{true}/N_\textrm{sat}}, \label{eq:sat}
\end{equation}
where $N_\textrm{meas}$ is the measured number of ions, $N_\textrm{true}$ is the unsaturated number of ions, and $N_\textrm{sat}$ is the saturation number, and assuming the the number of ions $N_\textrm{other}$ in states other than $32S_{1/2}$ and $31P_{3/2}$ is a fraction $\alpha$ of the total number of ions, the curve in Fig. \ref{fig:sat}a can be fit to:
\begin{equation}
    (N_{\ket{32S}} + N_{\ket{31P}})/N_\textrm{other} = \frac{(N_\textrm{sat}-(N_{\ket{32S}} + N_{\ket{31P}}))\alpha }{N_\textrm{sat}(1-\alpha)} 
    \label{eq:satfit}
\end{equation}
to extract the saturation number. The fit at $V_\textrm{det}$~=~1450~V yields $N_\textrm{sat} = 160 \pm 2$~mV$\cdot\mu$s. The inverse of Eq. \ref{eq:sat} is then taken to un-saturate the measured ion numbers:
\begin{equation}
    N_\textrm{true} = \frac{N_\textrm{meas}N_\textrm{sat}}{N_\textrm{sat} - N_\textrm{meas}}. 
    \label{eq:satcorr}
\end{equation}

The choice of a detector voltage of 1450V is made in order to keep the value of the maximum peak used in the measurement well below saturation, while maximizing the gain of the detector to increase statistical sensitivity. This choice is motivated by the measurements in Fig. \ref{fig:sat}b.

\begin{figure}[h]
a)
\includegraphics[trim = 0 -1.1cm 0 0, clip, width = .19\textwidth,valign = b]{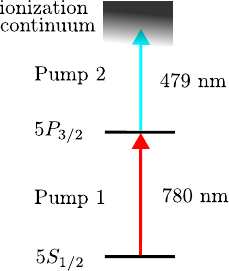}
b) \includegraphics[width = .375\textwidth, valign = b]{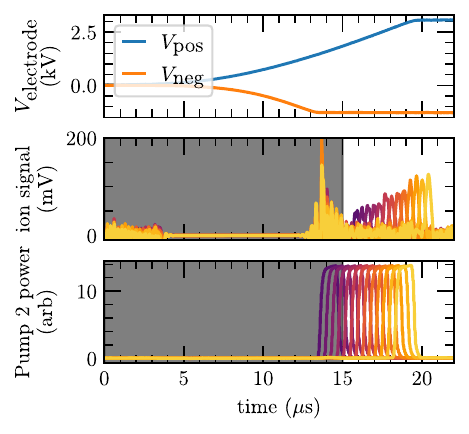}
c)
\includegraphics[width = .355\textwidth, valign = b]{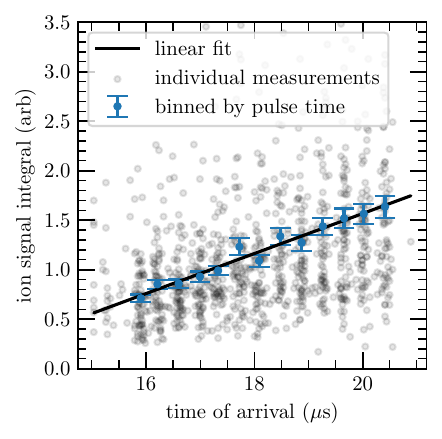}
\caption{Calibration of the detector gain as a function of time-of-arrival given the electric field ramp. a) Energy level diagram for the creation of ions on demand. b) Ions are created at different times in the electric field ramp (encapsulated by the Pump 2 power profile), leading to ions arriving at the detector at different times with different gains. The ion signal is unsaturated according to Eq. \ref{eq:satcorr}. c) For each pulse time, the time of arrival is determined from the peak of the ion signal, and the counted ion number is determined by integrating the ion signal. The values are then binned by the programmed ionization pulse time and fit to a first order polynomial.}
\label{fig:direct_ion}
\end{figure}

In addition, the ion detection efficiency can vary with the kinetic energy of the incident ion. Because the electrodes are being ramped during the detection, ions that ionize later will have a larger kinetic energy upon arrival compared to those that ionize earlier, and thus are detected with higher efficiency.  This effect can be measured by creating on-demand ions with direct photo-ionization.  Ions are created on-demand by pulsing the pump lasers in the scheme shown in Figure \ref{fig:direct_ion}a.

By generating ions on demand at various times throughout the ramp of the electric field, we can simulate Rydberg atoms getting field-ionized at that time in the ramp, and thus compare the measured signal on the channeltron to infer the relative gain for ions that are ionized at that time.  The number of ions created should not be effected by the ramp as the $5P_{3/2}$ is insensitive to electric fields.

The kinetic-energy dependent gain and therefore the time-of-arrival dependent gain $\mathcal{G}$ is fit to a first order expansion in the time-of-arrival:
\begin{equation}
    \mathcal{G} \propto \underbrace{(2.0\pm0.1) \times 10^5 \,\textrm{s}^{-1}}_{\equiv  \mathcal{A}} \cdot  t_\textrm{arrival} + \underbrace{(-2.5 \pm 0.2)\times 10^{3}}_{\equiv \mathcal{B}}. \label{eq:kegain}
\end{equation}
We can then account for this by dividing the measured ion signal by $\mathcal{G}(t_\textrm{arrival})$ before integrating the peaks to find the state populations.
\subsection{Uncertainty budget calculations}

To relate systematic errors to uncertainty in our temperature measurement, we must make some approximations. Consider the transition from the initially populated $32S_{1/2}$ state to the $32P_{3/2}$ state. At a transition frequency of 130~GHz and a temperature of 300~K, $\hbar \omega << k_B T$ and we can approximate the stimulated transition rate as \cite{Norrgard_2021}
\begin{equation}
   \Gamma_{32S\rightarrow 32P} \approx \frac{\mu^2 \omega^2}{3\epsilon_0 \pi \hbar^2 c^3} k_B T,
\end{equation}
which is linear in temperature. For small $t_\textrm{BBR}$, the ratio $\mathcal{R}$ can be approximated with a Taylor expansion with respect to time of the solution to the differential equation presented in the main text:
\begin{equation}
    \mathcal{R} = \Gamma_{32S\rightarrow 32P} \,t_\textrm{BBR} +\mathcal{O}(t_\textrm{BBR}^2).
\end{equation}
Thus to first order, the measured temperature goes as:
\def\approxprop{%
  \def\p{%
    \setbox0=\vbox{\hbox{$\propto$}}%
    \ht0=0.6ex \box0 }%
  \def\s{%
    \vbox{\hbox{$\sim$}}%
  }%
  \mathrel{\raisebox{0.7ex}{%
      \mbox{$\underset{\s}{\p}$}%
    }}%
}
\begin{equation}
    T_\textrm{meas} \approxprop \left(\frac{\mathcal{R}}{t_\textrm{BBR}} \right)\frac{1}{\mu^2}.
\end{equation}
From this we can take derivatives to find the associated temperature uncertainty $\sigma_T$ contribution from the uncertainty in each parameter:
\begin{table}[h!]
\begin{tabular}{l|c|r}
parameter & symbol & $\left(\frac{\sigma_T}{T}\right)^2$ contribution \\
\hline
blackbody evolution time & $t_\textrm{BBR}$ & $\left(\frac{\sigma_{t_\textrm{BBR}}}{t_\textrm{BBR}}\right)^2$\\
ratio of ion signal peak integrals & $\mathcal{R}$ & $\left(\frac{\sigma_{\mathcal{R}}}{\mathcal{R}}\right)^2$\\
measured integral of $32P_{3/2}$ peak & $N_{\ket{32P_{3/2}}}$ &  $\left(\frac{\sigma_{N_{\ket{32P_{3/2}}}}}{N_{\ket{32P_{3/2}}}}\right)^2$\\
measured integral of combined $32S_{1/2}$ and $31P_{3/2}$ peak & $N_{\ket{32S_{1/2}} + \ket{31P_{3/2}}}$ &  $\left(\frac{\sigma_{N_{\ket{32S_{1/2}} + \ket{31P_{3/2}}}}}{N_{\ket{32S_{1/2}} + \ket{31P_{3/2}}}}\right)^2$\\
\end{tabular}.
\end{table}

The fractional contribution of systematics to these parameters can be found by comparing their effect to typical parameters in the $t_\textrm{BBR}$ = 18-27 $\mu$s measurement window:
\begin{equation}
\begin{cases}
N_{\ket{32S} + \ket{31P}} &\approx 17\textrm{mV}\cdot \mu\textrm{s}\\
N_{\ket{32P}} &\approx 2.0 \textrm{mV}\cdot \mu\textrm{s}\\
t_\textrm{BBR} &\approx 23~\mu \textrm{s}
\end{cases}.
\end{equation}

\begin{figure}[h]
a)\includegraphics[height  = .26\linewidth]{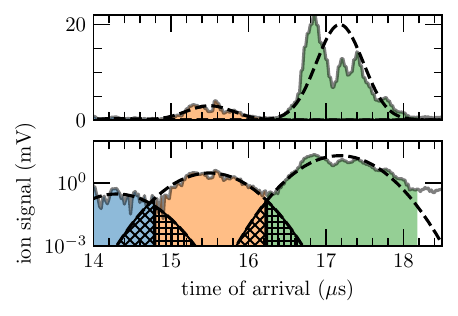}b)\includegraphics[height  = .26\linewidth]{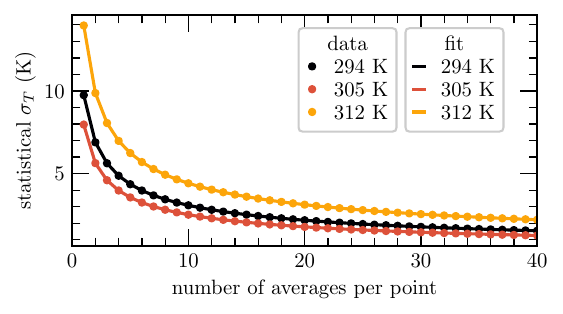}
\caption{Uncertainty characterization for a primary temperature measurement. a) Overlap of the peaks is determined by approximating them as normal distributions and integrating the area of overlap. The approximate normal distributions are drawn as dashed lines and compared to measured values on a linear (top) and logarithmic (bottom) scale, and the resulting overlap regions are represented with hatched areas.  b) The statistical sensitivity of the temperature measurement is found by fitting the uncertainty $\sigma_T$ vs the number of averages per point $N_\textrm{avg}$ according to Eq. \ref{eq:statfit}. The fits for each temperature are represented with solid lines, while the data is shown as points with error bars, which are too small to resolve.}
\label{fig:uncertainty}
\end{figure}

One possible source of uncertainty is the overlap of peaks on the ion time-of-flight signal. This can be characterized by approximating the peaks as normal distributions and numerically integrating the distributions' expected overlap, as demonstrated in Fig. \ref{fig:uncertainty}a. The largest contribution to uncertainty due to overlap of the peaks on the ion time-of-flight signal comes from the $(32S + 31P)$ peak leaking into the $32P$ peak. This is estimated to contribute 9$\pm$2~$\mu$V$\cdot$$\mu$s to this peak for typical operation, resulting in a fractional contribution to $\mathcal{R}$ and therefore to $T_\textrm{meas}$ of 0.005$\pm$0.001.

Another source of uncertainty is the determination of the initial time for $t_\textrm{BBR}$. That is, the finite times of state pumping and the ionization ramp require an effective treatment of $t_\textrm{BBR}$ where we allow for a small finite offset. The offset will be roughly 13.5~$\mu$s, which is found by measuring the time between pulsing the pump lasers and the ionized electrons arriving at the detector. However, this offset can be attained more precisely by optimizing agreement between theory and experiment as in Fig. 3 of the main text. Note that the shape and time of maximization of the $32P$ state is not particularly sensitive to temperature (only the magnitude is highly temperature-sensitive),  so performing this calibration does \textit{not} equate to calibrating the system via a known temperature. The fit   yields a $t_\textrm{BBR}$ equal to the time between the end of pumping and the start of the ionization ramp plus 13.97~$\mu$s. The uncertainty of this fit is 0.05~$\mu$s, so measuring at around $t_\textrm{BBR} = 23$~$\mu$s (measuring over a 40\% spread) gives a fractional $t_\textrm{BBR}$ uncertainty and thus a fractional temperature uncertainty of 0.002.

A third source of uncertainty is non-linearity of the channeltron detector due to saturation effects. This mainly affects the $(32S+31P)$ peak. The error introduced by using Eq. \ref{eq:satcorr} due to the uncertainty in $N_\textrm{sat}$ evaluates to a fractional uncertainty of 0.002.

In addition, the calibrated $t_\textrm{arrival}$-dependent gain $\mathcal{G}$ of the detector described in Eq. \ref{eq:kegain} contains associated uncertainty. The effect of the measurement on the ratio $\mathcal{R}$ can be approximated as a scaling by the applied corrections:
\begin{equation}
    \mathcal{R}\propto\frac{1/\mathcal{G}(t_\textrm{arrival} = 15.5 \,\mu\textrm{s})}{1/\mathcal{G}(t_\textrm{arrival} = 17 \,\mu\textrm{s})} = \frac{\mathcal{A}(17\,\mu\textrm{s}) + \mathcal{B}}{\mathcal{A}(15.5\,\mu\textrm{s}) + \mathcal{B}}.
\end{equation}
The resulting fractional uncertainty can then be written as
\begin{equation}
\frac{\sigma_T}{T} = \frac{\sigma_\mathcal{R}}{\mathcal{R}} = 
\frac{\sqrt{\left(\frac{\partial}{\partial \mathcal{A}}\left(\frac{\mathcal{A}(17\,\mu\textrm{s}) + \mathcal{B}}{\mathcal{A}(15.5\,\mu\textrm{s}) + \mathcal{B}}\right)\right)^2\sigma_\mathcal{A}^2 + \left(\frac{\partial}{\partial \mathcal{B}}\left(\frac{\mathcal{A}(17\,\mu\textrm{s}) + \mathcal{B}}{\mathcal{A}(15.5\,\mu\textrm{s}) + \mathcal{B}}\right)\right)^2\sigma_\mathcal{B}^2}}{\left(\frac{\mathcal{A}(17\,\mu\textrm{s}) + \mathcal{B}}{\mathcal{A}(15.5\,\mu\textrm{s}) + \mathcal{B}}\right)},
\end{equation}
where $\sigma_\mathcal{A}$ and $\sigma_\mathcal{B}$ are the uncertainties of the linear fit coefficients of the gain. This evaluates to a fractional uncertainty of $1\times 10^{-5}$.

Artefacts from the pulsing of the ionization field can be seen due to capacitive coupling to the channeltron detectors. The main artefact comes from ringing contaminating the $32P_{3/2}$ peak, with an estimated integral of $5\pm2$~$\mu \textrm{V} \cdot  \mu\textrm{s}$, resulting in a fractional contribution to $\mathcal{R}$ and thus to $T_\textrm{meas}$ of $0.003\pm 0.001$.

Finally, the statistical sensitivity is found by fitting the statistical uncertainty of the primary temperature measurement as a function of the number of averages in Fig. \ref{fig:uncertainty}b. The uncertainty is fit to:
\begin{equation}
\sigma_T = \mathcal{S}(\sigma_T/T) \cdot T_\textrm{env} \frac{1}{\sqrt{\tau_\textrm{meas}}} =   \mathcal{S}(\sigma_T/T) \cdot T_\textrm{env} \frac{1}{\sqrt{N_\textrm{points}N_\textrm{avg}\tau_\textrm{shot}}}  =   \left(\mathcal{S}(\sigma_T/T) \cdot T_\textrm{env} \frac{1}{\sqrt{N_\textrm{points}\tau_\textrm{shot}}} \right) \frac{1}{\sqrt{N_\textrm{avg}}},
\label{eq:statfit}
\end{equation}
where $\mathcal{S}(\sigma_T/T)$ is the statistical sensitivity in Hz$^{-1/2}$, $T_\textrm{env}$ is the temperature at which the measurement is made, $\tau_\textrm{meas}$ is the total measurement time, $\tau_\textrm{shot}$ is the measurement time of an individual shot including dead time equal to 354~ms, $N_\textrm{points}$ is the number of points in the temperature measurement equal to 20, and $N_\textrm{avg}$ is the number of averages per point. The fit sensitivities for each measurement are:

\begin{table}[h!]
\begin{tabular}{c|c}
$T_\textrm{env}$ (K) & $\mathcal{S}(\sigma_T/T)$ (Hz$^{-1/2}$)\\
\hline
   294  &   0.088\\
   305  &  0.068\\
   312 & 0.111\\
\end{tabular},
\end{table}
\noindent yielding a mean sensitivity of $\mathcal{S}(\sigma_T/T)$ of 0.086 $\pm$ 0.009 Hz$^{-1/2}$.

\bibliography{bibliography}